# Effect of minimal length uncertainty on neutrino oscillation

Fidele J. Twagirayezu[1]

[1]University of California, Los Angeles, Department of Physics and Astronomy, Los Angeles, CA, 90095, USA

## Abstract:

In this paper, we study the effect of the minimal length on neutrino oscillation in a static magnetic field. In the framework of the generalized uncertainty principle, we reformulate the Hamiltonian for a relativistic neutrino moving in a magnetic field oriented along the z-direction of Cartesian coordinates. Using the modified energy spectrum, we obtain the oscillation probability for different neutrino flavors. In addition, we obtain the energy differences for the neutrino-mass eigenstates. We find that the energy and energy difference depend on the minimal length parameter $\alpha$, and the energy difference becomes independent of $\alpha$ when the magnetic field is not present. In addition, we find that the modified probability of oscillation differs from the usual probability of oscillation if a magnetic field is present. Using the current experimental result, we estimate the upper bound on the deformation parameter and the minimal length, and find that the upper bound on the minimal length scale is less than the electroweak scale. If the minimal length is at Planck scale, the minimal length formalism leads to the same result as a quantum theory of gravity with an $SU(2)_L \otimes U(1)$ effective invariant dimension-5 Lagrangian including neutrino and Higgs fields.

## Keywords:

Quantum gravity, generalized uncertainty principle, minimal length, neutrino oscillation, early universe

1. ## Introduction:

It has been found that neutrinos oscillate between different flavors as they move throughout space. In the Standard model neutrinos are massless, but in the theory of grand unification also known as GUT, neutrinos are assumed to obtain mass by see-saw mechanism. Neutrinos play an important role in cosmology and astroparticle physics; for instance, it is strongly believed that neutrino oscillation in extreme conditions might have played an essential role in the generation of lepton asymmetry in the early universe; during the electroweak phase transition in the early universe, leptons obtain mass by electroweak symmetry breaking and then a mass-difference between neutrino and antineutrino would create a difference in the chemical potential, leading to lepton asymmetry. Because gravity was strong in the early universe, it is interesting to study the effect of gravity on the neutrino oscillation in the early universe, which would have an impact on the understanding of cosmic microwave background radiation. As already mentioned in the beginning the mass of neutrinos are believed to be generated by see-saw mechanics, but contributions from other sources are likely to exist. The most probable sources are gravitational interactions. For instance starting from the Lagrangian density, the relevant effective gravitational interaction of neutrinos with Higgs fields can be expressed as an invariant dimension 5-operator for the spinor $SU(2)_L$ iso-doublets and the scalar one [1, 2]:

$$\mathcal{L}_{\text{grav}} = \frac{\lambda_{\alpha\beta}}{M_p}(\psi_{Aa\alpha}\varepsilon_{AC}h_C)\tilde{C}_{ab}^{-1}(\psi_{Bb\beta}\varepsilon_{BD}h_D) + h.c \tag{1}$$

where $\alpha, \beta$ are the flavor states, $\psi_\alpha = (v_\alpha, l_\alpha)$ is the lepton doublet, $h = (h^+, h^\circ)$ is the Higgs doublet, and $M_p \approx 10^{19}\,GeV$ is the Planck mass. $\lambda_{\alpha\beta}$ is a matrix in a flavour space with elements $O(1)$. $\tilde{C}_{ab}$ is the charge conjugation matrix with Lorentz indices $a, b = 1, 2, 3, 4$. $A, B, C, D = 1, 2$ are the isospin indices and $\varepsilon = i\sigma_2$ where $\sigma_m, m = 1, 2, 3$ are the Pauli matrices. After the spontaneous symmetry breaking, Eq. (1) generates an additional term of neutrino mass of the form [1, 2]

$$\mathcal{L}_{\text{mass}} = \frac{v^2}{M_p}\lambda_{\alpha\beta}v_\alpha\tilde{C}^{-1}v_\beta \tag{2}$$

where $v$ is the vacuum expectation value of the electroweak symmetry breaking. Quantum gravity effects can lead to CPT violation. From Eq. (2) the CPT violating mass is $v^2/M_p$. The corresponding mass matrix can be expressed as [1, 2]

$$M_{CPT} = \mu\lambda_{\alpha\beta} \tag{3}$$

where $\mu$ is a parameter with a mass dimension. Assuming that the gravitational interaction is insensitive to neutrino flavors, this means that $\lambda_{\alpha\beta}$ is independent of $\alpha, \beta$ then the Planck scale contribution to the neutrino [1, 2]

$$\mu\lambda = \mu\begin{pmatrix} 1 & 1 & 1 \\ 1 & 1 & 1 \\ 1 & 1 & 1 \end{pmatrix} \tag{4}$$

For clarity, $\lambda_{\alpha\beta}$ becomes $\lambda$ in quantum gravity. For $v = 174\,GeV$, the value for the scale $\mu$ is obtained in the following way [1, 2]

$$\mu = v^2/M_p \approx 2.5 \times 10^{-6} \,\text{GeV} \tag{5}$$

However, other interactions of the form (1) with $M_X$ less that $M_p$ may exist, for example, from string compactification, non-perturbative dynamics or simply higher GUT scales. These interactions do necessarily contain $\alpha, \beta$ indices, thus this leads to a perturbation scale $\mu = v^2/M_X$ and a structure modification of the perturbation matrix $\lambda_{\alpha\beta}$. The operators with the scale $M_{GUT} \leq M_X < M_p$ are believed to produce much stronger effects. Here, $M_{GUT}$ is the supersymmetric unification scale, $M_{GUT} \sim 2 \times 10^{19} \,\text{GeV}$.

In addition, the implementation of minimal length allows studying quantum gravitational effects in physical phenomena. The first person who realized the importance of minimal length for the description of high-energy phenomena was Heisenberg [3,4]. The consideration of minimal length leads to the modified Heisenberg uncertainty principle known as the generalized or gravitational uncertainty principle (GUP) [5,6,7,8,9,10,11,12,13], which can be expressed as follows

$$\Delta X \Delta P \geq \frac{\hbar}{2}\left[1 + \phi(\Delta P)\right] \tag{6}$$

where $\phi$ is a function that represents a correction to the usual Heisenberg uncertainty principle. In terms of the deformation parameters, Eq. (6) can be written as follows [14,15]

$$\Delta X^i \Delta P^i \geq \frac{\hbar}{2}\left[1 + (\alpha D + \alpha')(\Delta P^i)^2 + \alpha \sum_{k=1}^{D}\langle P^k \rangle^2 + \alpha'\langle P^i \rangle^2\right] \tag{7}$$

where $i = 1, 2, ..., D$ and $\alpha, \alpha'$ are non-negative deformation parameters with the same dimensions. The minimal length can be expressed as follows [16-18]

$$(\Delta X^{i \in \{1,2,...,D\}})_{MIN} = \hbar\sqrt{(D\alpha + \alpha')} \tag{8}$$

In the case where $\alpha' = 2\alpha$, the modified position and momentum operators that satisfy Eq. (7) can be written as follows [19]

$$\begin{aligned} X^i &= x^i \\ P^i &= p^i(1 + \alpha \mathbf{p}^2) \end{aligned} \tag{9}$$

where $x^i$, and $p^i \doteq -i\hbar\partial_i$ are position and momentum operators in the usual Heisenberg algebra.

This article is organized as follows: In Sec. 2, we derive the modified relativistic energy spectrum for the neutrino moving in the static magnetic field along the $z$-direction in Cartesian coordinates. In Sec. 3, we consider two neutrino flavors and obtain the probability of neutrino oscillation and the modified energy difference spectrum. In Sec. 4, we summarize the content of our article.

## 2. Relativistic dispersion relation with a minimal length

The usual dynamics for a relativistic particle of charge $e$, mass $m$ and magnetic moment $\mu_0 = \mp \mu$ moving in a magnetic potential $\mathbf{A}$ can be described by the following Hamiltonian

$$H = c\left\{\left(\mathbf{p} - \frac{e}{c}\mathbf{A}\right)^2 + \frac{(mc^2)^2}{c^2}\right\}^{1/2} \pm \mu(\nabla \times \mathbf{A}) \tag{10}$$

The first term in Eq. (10) is just the Hamiltonian for a spinless relativistic particle interacting with an external magnetic potential [20], and the second term in Eq. (10) is the Hamiltonian arising from the spin interaction with an external magnetic potential. If we consider Eq. (9) and Eq. (10), the modified Hamiltonian can be expressed as

$$H_{\text{MIN}} = c\left\{\left(\mathbf{p} - \frac{e}{c}\mathbf{A}\right)^2\left[1 + \alpha\left(\mathbf{p} - \frac{e}{c}\mathbf{A}\right)^2\right]^2 + \frac{(mc^2)^2}{c^2}\right\}^{1/2} \pm \mu\left(1 + \alpha \mathbf{p}^2\right)(\nabla \times \mathbf{A}) \tag{11}$$

where subscript (MIN) indicates that a minimal length scale has been implemented. The modified energy spectrum can be expressed as follows

$$\left(E_\Theta^\pm\right)_{\text{MIN}} = c\left\{\left(\mathbf{p} - \frac{e}{c}\mathbf{A}\right)^2\left[1 + \alpha\left(\mathbf{p} - \frac{e}{c}\mathbf{A}\right)^2\right]^2 + \frac{(mc^2)^2}{c^2}\right\}^{1/2} \pm \mu^B\left(1 + \alpha \mathbf{p}^2\right)(\nabla \times \mathbf{A}) \tag{12}$$

Consider Cartesian coordinates. If we choose the magnetic vector potential $\mathbf{A}$ to be oriented along the $x$-direction so that the magnetic field $\mathbf{B}$ is along the $z$-direction, then Eq. (12) is expressed as follows

$$\left(E_\Theta^\pm\right)_{\text{MIN}} = c\left\{\left[\left(p_x - \frac{e}{c}A_x\right)^2 + p_y^2 + p_z^2\right]\left[1 + \alpha\left(p_x - \frac{e}{c}A_x\right)^2 + \alpha p_y^2 + \alpha p_z^2\right]^2 + \frac{(mc^2)^2}{c^2}\right\}^{1/2}$$
$$\pm \mu^B\left(1 + \alpha p^2\right)B \tag{13}$$

If we expand to first-order in $\alpha$ the expression in the second bracket of Eq. (13), then we can write

$$\left(E_\Theta^\pm\right)_{\text{MIN}} = c\left\{\left[\left(p_x - \frac{e}{c}A_x\right)^2 + p_y^2 + p_z^2\right]\left[1 + 2\alpha\left(p_x - \frac{e}{c}A_x\right)^2 + 2\alpha p_y^2 + 2\alpha p_z^2\right] + \frac{(mc^2)^2}{c^2}\right\}^{1/2} \quad (14)$$
$$\pm \mu^B\left(1 + \alpha p^2\right)B$$

In the weak magnetic field, Eq. (14) becomes

$$\left(E_\Theta^\pm\right)_{\text{MIN}} = c\left\{\left(p^2 - 2\frac{e}{c}p_xA_x\right)\left[1 + 2\alpha\left(p^2 - 2\frac{e}{c}p_xA_x\right)\right] + \frac{(mc^2)^2}{c^2}\right\}^{1/2} \pm \mu^B\left(1 + \alpha p^2\right)B$$

$$= c\left\{\left(p^2 - 2\frac{e}{c}p_xA_x\right) + 2\alpha p^2 c^2\left(p^2 - 4\frac{e}{c}p_xA_x\right) + \frac{(mc^2)^2}{c^2}\right\}^{1/2} \pm \mu^B\left(1 + \alpha p^2\right)B \quad (15)$$

In the highly relativistic regime, Eq. (15) can be expressed as

$$\left(E_\Theta^\pm\right)_{\text{MIN}} = c\left\{c\left(p^2 - 2\frac{e}{c}p_xA_x\right) + 2\alpha p^2\left(p^2 - 4\frac{e}{c}p_xA_x\right) + \frac{(mc^2)^2}{c^2}\right\}^{1/2} \pm \mu^B\left(1 + \alpha p^2\right)B$$

$$\approx c\left\{p + \frac{(mc^2)^2}{2pc^2} - 2\left(\frac{e}{c}\right)\frac{p_xA_x}{p} + 2\alpha p\left(p^2 - 4\frac{e}{c}p_xA_x\right)\right\} \pm \mu^B\left(1 + \alpha p^2\right)B \quad (16)$$

If we choose the following gauge condition, $A_x = -By/2$, Eq. (16) becomes

$$\left(E_\Theta^\pm\right)_{\text{MIN}} \approx cp + \frac{(mc^2)^2}{2pc} + cB\left(\frac{e}{c}\right)\frac{p_xy}{p} + 2\alpha pc\left(p^2 + 2B\frac{e}{c}p_xy\right) \pm \mu^B\left(1 + \alpha p^2\right)B$$

$$\approx E_\Theta + c\left(\frac{e}{c}\right)\frac{BL_z}{p} + 2\alpha pc\left(p^2 + 2\frac{e}{c}BL_z\right) \pm \mu^B\left(1 + \alpha p^2\right)B \quad (17)$$

where $L_z$ is the component of angular momentum $\mathbf{L}$ along the $z$-direction. $E_\Theta$ is the usual energy without the contribution from the magnetic interaction. Notice that the above results can easily be applicable to neutrinos if we consider, $e \to 0$.

### 3. Modified probability for neutrino oscillation

Considering two types of neutrinos, the relation between the weak interaction eigenstates $|v_e\rangle$, $|v_\mu\rangle$, and mass eigenstates $|v_1\rangle$, $|v_2\rangle$ is

$$|v_e\rangle = \cos\frac{\theta}{2}|v_1\rangle + \sin\frac{\theta}{2}|v_2\rangle$$
$$|v_\mu\rangle = \sin\frac{\theta}{2}|v_1\rangle - \cos\frac{\theta}{2}|v_2\rangle \tag{18}$$

where $\theta/2$ is the mixing angle.

Assume that an electron neutrino is created in some weak interaction process and propagates through a modified static magnetic field $(1+\alpha p^2)B$ [21] to a detector. We are interested in knowing the modified probability that a muon neutrino is detected, which indicates neutrino flavor mixing. The initial state is

$$|\psi(0)\rangle_{MIN} = |v_e\rangle$$
$$= \cos\frac{\theta}{2}|v_1\rangle + \sin\frac{\theta}{2}|v_2\rangle = |\psi(0)\rangle \tag{19}$$

The Schrodinger time evolution for Eq. (19) is

$$|\psi(t)\rangle_{MIN} = \cos\frac{\theta}{2}e^{-i(E_\Theta^\pm)_{MIN}t/\hbar}|v_1\rangle + \sin\frac{\theta}{2}e^{-i(E_\Theta^\pm)_{MIN}t/\hbar}|v_2\rangle \tag{20}$$

Because the $(\pm)$ corresponds to two different eigenvalues, Eq. (20) can be expressed as follows

$$|\psi(t)\rangle_{MIN} = \cos\frac{\theta}{2}\left(c_1^- e^{-i(E_1^-)_{MIN}t/\hbar}|v_1^-\rangle + c_1^+ e^{-i(E_1^+)_{MIN}t/\hbar}|v_1^+\rangle\right)$$
$$+ \sin\frac{\theta}{2}\left(c_2^- e^{-i(E_2^-)_{MIN}t/\hbar}|v_2^-\rangle + c_2^+ e^{-i(E_2^+)_{MIN}t/\hbar}|v_2^+\rangle\right) \tag{21}$$

The modified probability of a neutrino oscillation is expressed as follows

$$\left(\wp_{v_e \to v_\mu}\right)_{MIN} = \left(\langle v_\mu|\psi(t)\rangle_{MIN}\right)\left(\langle v_\mu|\psi(t)\rangle_{MIN}\right)^*$$
$$= \left|\langle v_\mu|\left\{\cos\frac{\theta}{2}\left(c_1^- e^{-i(E_1^-)_{MIN}t/\hbar}|v_1^-\rangle + c_1^+ e^{-i(E_1^+)_{MIN}t/\hbar}|v_1^+\rangle\right)\right.\right.$$
$$\left.\left. + \sin\frac{\theta}{2}\left(c_2^- e^{-i(E_2^-)_{MIN}t/\hbar}|v_2^-\rangle + c_2^+ e^{-i(E_2^+)_{MIN}t/\hbar}|v_2^+\rangle\right)\right\}\right|^2 \tag{22}$$

Inserting the muon state from Eq. (18) into Eq. (22) leads to

$$\left(\wp_{\nu_e \to \nu_\mu}\right)_{MIN} = \sin^2\left(\frac{\theta}{2}\right)\cos^2\left(\frac{\theta}{2}\right)\Big|\Big\{\left|c_1^-\right|^2 e^{-i(E_1^-)_{MIN} t/\hbar} + \left|c_1^+\right|^2 e^{-i(E_1^+)_{MIN} t/\hbar}$$
$$-\left|c_2^-\right|^2 e^{-i(E_2^-)_{MIN} t/\hbar} - \left|c_2^+\right|^2 e^{-i(E_2^+)_{MIN} t/\hbar}\Big\}\Big|^2 \quad (23)$$

The expansion of Eq. (23) leads to

$$\left(\wp_{\nu_e \to \nu_\mu}\right)_{MIN} =$$
$$\sin^2\left(\frac{\theta}{2}\right)\cos^2\left(\frac{\theta}{2}\right)\Big[\left|c_1^-\right|^2\left|c_1^-\right|^2 + \left|c_1^+\right|^2\left|c_1^+\right|^2 + \left|c_2^-\right|^2\left|c_2^-\right|^2 + \left|c_2^+\right|^2\left|c_2^+\right|^2$$
$$+2\left|c_1^-\right|^2\left|c_1^+\right|^2\cos\left\{\left((E_1^+)_{MIN} - (E_1^-)_{MIN}\right)t/\hbar\right\} + 2\left|c_2^-\right|^2\left|c_2^+\right|^2\cos\left\{\left((E_2^+)_{MIN} - (E_2^-)_{MIN}\right)t/\hbar\right\} \quad (24)$$
$$-2\left|c_1^+\right|^2\left|c_2^+\right|^2\cos\left\{\left((E_1^+)_{MIN} - (E_2^+)_{MIN}\right)t/\hbar\right\} - 2\left|c_1^-\right|^2\left|c_2^-\right|^2\cos\left\{\left((E_1^-)_{MIN} - (E_2^-)_{MIN}\right)t/\hbar\right\}$$
$$-2\left|c_1^+\right|^2\left|c_2^-\right|^2\cos\left\{\left((E_1^+)_{MIN} - (E_2^-)_{MIN}\right)t/\hbar\right\} - 2\left|c_1^-\right|^2\left|c_2^+\right|^2\cos\left\{\left((E_1^-)_{MIN} - (E_2^+)_{MIN}\right)t/\hbar\right\}\Big]$$

Using Eq. (17), expressions for energy differences in Eq. (24) are

$$\left(E_1^+\right)_{MIN} - \left(E_1^-\right)_{MIN} = 2(1+\alpha p^2)\mu_1^B B$$
$$\left(E_2^+\right)_{MIN} - \left(E_2^-\right)_{MIN} = 2(1+\alpha p^2)\mu_2^B B$$
$$\left(E_1^+\right)_{MIN} - \left(E_2^+\right)_{MIN} = \frac{m_1^2 - m_2^2}{2p}c^3 + (1+\alpha p^2)\left(\mu_1^B - \mu_2^B\right)B$$
$$\left(E_1^-\right)_{MIN} - \left(E_2^-\right)_{MIN} = \frac{m_1^2 - m_2^2}{2p}c^3 - (1+\alpha p^2)\left(\mu_1^B - \mu_2^B\right)B \quad (25)$$
$$\left(E_1^-\right)_{MIN} - \left(E_2^+\right)_{MIN} = \frac{m_1^2 - m_2^2}{2p}c^3 - (1+\alpha p^2)\left(\mu_1^B + \mu_2^B\right)B$$
$$\left(E_1^+\right)_{MIN} - \left(E_2^-\right)_{MIN} = \frac{m_1^2 - m_2^2}{2p}c^3 + (1+\alpha p^2)\left(\mu_1^B + \mu_2^B\right)B$$

We realize that if the magnetic field is off, the modified energy difference would be independent of $\alpha$, but the modified energy would still depend on $\alpha$. Since Eq. (24) depends on the modified energy difference; thus, the modified probability of oscillation would be different from the usual probability of oscillation if a magnetic field is present. From Eq. (25), we can write the following expression

$$\frac{c^3}{2p}\left(m_1^2 - m_2^2\right)_{MIN} + \left(\mu_1^B + \mu_2^B\right)B = \frac{c^3}{2p}\left(m_1^2 - m_2^2\right) + (1+\alpha p^2)\left(\mu_1^B + \mu_2^B\right)B \quad (26)$$

Let us define positive quantities, $(\Delta m^2) = (m_1^2 - m_2^2)$, and $(\Delta m^2)_{MIN} = (m_1^2 - m_2^2)_{MIN}$. Then, Eq. (26) becomes

$$(\Delta m^2)_{MIN} = (\Delta m^2) + 2\alpha \frac{p^3}{c^3}(\mu_1^B + \mu_2^B)B \tag{27}$$

Eq. (27) shows that even if neutrinos were massless, they would still have a mass in the magnetic field if gravity is taken into account. It is important to recall that the magnetic field is not completely zero in space or inside particle colliders; therefore, the only requirement for the above argument is quantum gravity. Because the minimal length can be written as $(\Delta X)_{MIN} = \tilde{\alpha} L_P$, where $\tilde{\alpha}$ is a dimensionless deformation parameter and $L_P$ is the Planck length, then using Eq. (8), we can rewrite Eq. (27) as follows

$$(\Delta m^2)_{MIN} = (\Delta m^2) + 2\tilde{\alpha}^2 \frac{p^3 L_P^2}{5\hbar^2 c^3}(\mu_1^B + \mu_2^B)B \tag{28}$$

If we consider the case in which the term in the $\tilde{\alpha}^2$ is smaller than the left-hand side of Eq. (28), we can write

$$(\Delta m^2)_{MIN} \geq 2\tilde{\alpha}^2 \frac{p^3 L_P^2}{5\hbar^2 c^3}(\mu_1^B + \mu_2^B)B \tag{29}$$

We realize that each correction term in Eqs. (27), (28), and (29) increases significantly with $p$ as compared to $B$.

If we consider the atmospheric magnetic field, $B \approx 10^{-2}$ Tesla, the neutrino magnetic moment of $\mu_1^B \approx \mu_2^B \approx 10^{-10}$ magneton, and the mass-difference of $\approx 10^{-6} (eV/c^2)^2$, then for an average neutrino energy $E \approx cp \approx 10^{10}$ eV

$$10^{-6} \geq \tilde{\alpha}^2 10^{-20}$$
$$\tilde{\alpha} \leq 10^7 \tag{30}$$

The upper bound on $\tilde{\alpha}$ is better than those which are found in Refs. [22], [24]. Using Eqs. (8), (30), the upper bound on the minimal length is

$$(\Delta X)_{MIN} = \tilde{\alpha} L_P \leq 10^{-28} m \tag{31}$$

Eq. (31) indicates that the upper bound on the minimal length is less than the electroweak length scale, $\approx 10^{-18} m$, and less than the upper bound found in Ref. [23] for a non-relativistic case.

For the atmospheric neutrino, if the minimal length is equal or very close to its upper bound then Eq. (30) shows that quantum gravitational effects would be measurable, but if the minimal length is at the Planck scale, $\tilde{\alpha}=1$, then Eq. (30) shows that quantum gravitational effects would be insignificant. Since Eqs. (27), (28), and (29) show that gravitational effects increase with $p$ then we can estimate the neutrino mass-difference for which quantum gravitational effects are observable if the minimal length is at the Planck scale. Let us assume that the magnetic field strength throughout space is close to the atmospheric magnetic field, $B \approx 10^{-2}$ Tesla. Now, consider a neutrino of an energy, $E \approx cp \geq 10^{15}$ eV, then using Eq. (29) we obtain

$$\left(\Delta m^2\right)_{MIN} \geq 2 \frac{E^3 L_p^2}{5\hbar^2 c^6}\left(2\tilde{\mu}B\right)$$
$$\left(\Delta m^2\right)_{MIN} \geq 2.0 \times 10^{-6} \text{ eV}^2/c^4$$
(32)

where $\tilde{\mu} = \mu_1^B \approx \mu_2^B$. Eq. (32) shows that quantum gravitational effects would be measureable for a neutrino energy of, $E \geq 10^{15}$ eV, for example, galactic or extra-galactic neutrinos. Since the CPT violation for neutrinos is expected to take place at high energy we can compare our high-energy result from Eq. (32) with the one which is obtained due to the CPT violating effect. The corrected mass-differences obtained using Eq. (1) are $\sim 10^{-3}$ eV$^2$ or $\sim 10^{-6}$ eV$^2/c^4$ [2], thus we find that, for $E = 10^{15}$ eV the minimal length effect and the CPT violation effect lead to the mass-difference correction term of almost the same magnitude.

**Summary**

The search for the final laws of nature also requires the probing of initial conditions of the universe, since gravity was strong in the very early universe, and neutrinos were among particles that inhabited the early universe, we thought that a study on the effects of gravity on some properties of neutrino oscillation was necessary. In our study, we found that the energy spectrum depends on the minimal length. The energy spectrum still depends on the minimal length if the magnetic field is not present. The probability of oscillation depends on the minimal length through the modified energy spectrum. The energy difference depends only on the minimal length if the magnetic field is present. Using the current experimental result, the upper bound on the minimal length was found to be less than the electroweak scale. In addition, the results about correction terms from this study continue to support the belief that quantum gravity effects become more important at higher energy scales.

**Acknowledgements**

F. T. acknowledges the Eugene Cota-Robles Fellowship. F. T. would like to thank the UCLA Graduate Division.